# 1.1 kW, 100 Hz room-temperature diode-pumped nanosecond laser by water immersion cooling

**Xinxing Lei,**[1,2,3] **Suyang Wang,**[1,2,3] **Zichao Wang,**[1,2,3] **Lei Huang,**[1,2,3*] **Qiang Liu,**[1,2,3*] **and Xing Fu**[1,2,3*]

[1]*Department of Precision Instrument, Tsinghua University, Beijing, China.*

[2]*State Key Laboratory of Precision Space-time Information Sensing Technology, Beijing, China.*

[3]*Key Laboratory of Photonic Control Technology (Tsinghua University), Ministry of Education, Beijing, China.*

*hl@tsinghua.edu.cn
*qiangliu@tsinghua.edu.cn
*fuxing@tsinghua.edu.cn

**We report a room-temperature diode-pumped solid-state Nd:YAG laser by water immersion cooling, which delivers a pulse energy of 11 J at the repetition rate of 100 Hz and the pulse duration of 7 ns, while the beam quality factor is 2.6 times the diffraction limit. To the best of our knowledge, this represents the highest performance achieved for room-temperature nanosecond lasers operating above 100 Hz, which demonstrates the great potentials of room-temperature immersion-cooled nanosecond active mirror lasers.**

Diode-pumped solid-state lasers (DPSSLs), as the new generation of high-performance lasers, are enabling a wide range of industrial applications and scientific investigations, such as laser shock peening, femtosecond laser pumping, and strong-field physics [1-3]. There are currently two main approaches to achieve high-energy DPSSLs with high repetition rates (e.g., >10 J per pulse at >10 Hz) in terms of gain materials: cryogenic Yb-doped gain medium and room-temperature Nd-doped gain medium. Based on cryogenically cooled Yb:YAG, the Bivoj/DiPOLE100 project has demonstrated outputs of 146 J at 10 Hz in 2021 [4] and 10 J at 100 Hz (operated for 45 minutes) in 2024 [5] using multi-slab amplifier configuration. In addition, Osaka University achieved 10 J at 100 Hz using a cryogenically cooled Yb:YAG active mirror amplifier (operated for 20 seconds) [6]. Compared with Nd:YAG, the cryogenically cooled Yb:YAG laser exhibits a lower quantum defect, a higher saturation fluence, and a reduced requirement for pump peak power. However, these lasers require cryogenic cooling, which increases the system complexity and footprint.

Alternatively, taking the advantage of room-temperature operation of Nd-doped gain medium, our group reported 10 J at 10 Hz in 2019 [7], and 100 J at 10 Hz in 2021 [8] using hybrid distributed active mirror amplifier chain with Nd:LuAG and Nd:YAG, while Jiang et al. realized 10 J at 50 Hz in 2022 using a Nd:YAG active mirror structure [9]. However, the previous active mirror geometry commonly adopts single-sided cooling [7-10] that the limited cooling area results in elevated temperatures within the gain medium, which severely limits further enhancement of repetition rate and average power. Additionally, the single-side cooling creates a temperature gradient that monotonically increases along the thickness of the gain medium, leading to significant deformation that degrade the spatial overlap and beam quality. In addition, the ELI/LLNL HAPLS system achieved a 97 J output at 3.3 Hz in 2017, employing a Nd-glass multi-slab amplifier architecture [11].

In this work, we report the stable room-temperature operation of a Nd:YAG laser obtaining an output energy of 11 J at the repetition rate of 100 Hz and the pulse duration of 7 ns, using an active mirror structure with double-sided immersion cooling by water, which has several marked features. First, by employing a symmetrical double-sided cooling of gain medium, the temperature gradient and thermal deformation are significantly reduced, compared with previous active-mirror system. Second, by stacking two or more active-mirror slabs into one gain module (i.e., a packaged unit containing the gain medium and its coolant structure), the footprint can be effectively diminished. Compared to the multi-slab configuration [4-6] that also employs stacked gain media, the active mirror geometry achieves higher gain due to its nature of round-trip energy extraction. Third, despite the extracting laser beam passes through a dozen of water layers that may affect the wavefront [12], we verify that the beam quality of 1.1 kW, 100 Hz room-temperature nanosecond laser can be controlled within 2.6 times diffraction limit, with the help of lenses and adaptive optics (AO) system.

The optical layout of our immersion-cooled active mirror laser is illustrated in Fig. 1. The seed beam is generated from a homemade Q-switched side-pumped Nd:YAG rod oscillator, capable of delivering the pulse energy of 200 mJ at the duration of 7 ns (with a near-Gaussian temporal profile) and the frequency of 100 Hz. After beam expansion, the beam was cropped by a square aperture to a 30 mm × 30 mm square beam, with an energy of 100 mJ, which then enters the amplifier. The amplifier consists of five immersion-cooled gain modules, first two of which employ polarization multiplexing for double-pass amplification, while the following three modules utilize single-pass amplification.

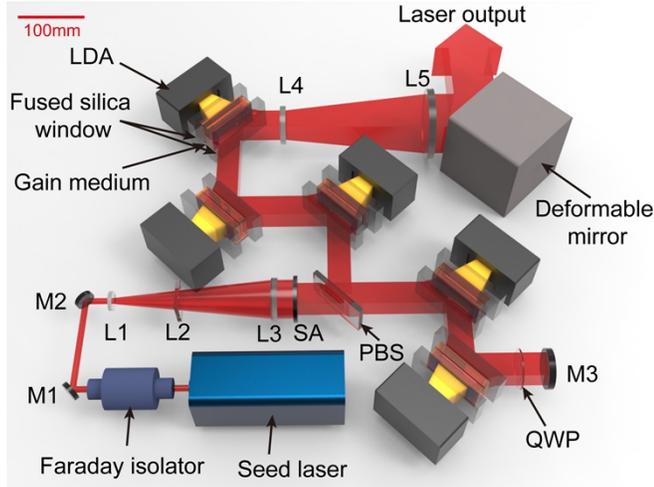

Fig. 1. Optical layout of the Nd:YAG laser system. M1, M2, M3, mirrors; L1, concave lens (f = -50 mm); L2, cylindrical concave lens (f = -1000 mm); L3, convex lens (f = 250 mm); L4, concave lens (f = -300 mm); L5, convex lens (f = 550 mm); SA, serrated aperture; PBS, dielectric-coated plate-type polarizing beam splitter; QWP, quarter-wave plate; LDA, laser diode array.

Figure 2 shows the schematic diagram of each gain module, having two Nd:YAG slab crystals with the dimension of 35 mm × 70 mm × 10 mm, and the doping concentrations of 0.2 at.% and 0.5 at.% respectively. To absorb the amplified spontaneous emission (ASE) light, 3-mm-thick samarium-doped YAG (Sm:YAG) ceramics are bonded to the outer sides of the gain medium. The Sm:YAG provide absorption at 1064 nm while being transparent to the 808 nm pump light. Within each gain modules, the two Nd:YAG slabs and two fused silica windows are placed in parallel, forming three flow channels of deionized water with the width of 0.4 mm for effective direct cooling of slabs. A 1064 nm high reflection (HR) coating is applied to the window surface where the extracting beam folds, while all other optical surfaces are coated with 1064 nm anti-reflection (AR) coating. The laser beam enters the gain module at an incidence angle of 45°, and extracts the stored gain along the V-shaped path. Each gain module is pumped from the single side by a laser diode array (LDA), which has the peak pump power of 60 kW at the duration of 300 μs and the central wavelength of 808 nm. The pump light is diverging along the slab thickness through a specially designed coupling system, to ensure an efficient spatial overlap with the V-shaped extracting beam path.

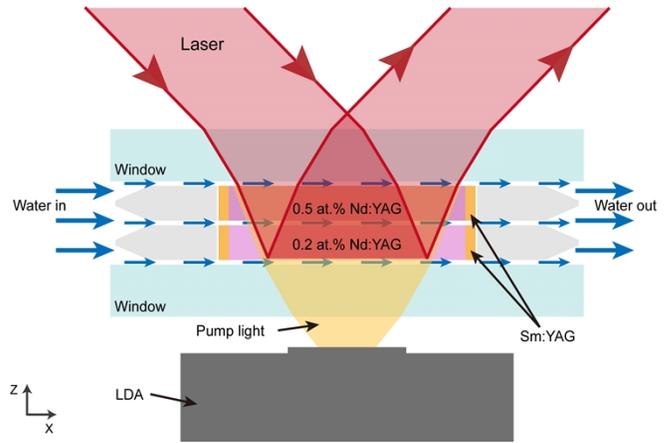

Fig. 2. Schematic diagram of the immersion-cooled active-mirror gain module

Temperature simulation has been performed using finite element software with key parameters set to match the experimental conditions: a cooling flow velocity of 5 m/s (resulting in a convective heat transfer coefficient of 10,000 W/(m²·K)) and a coolant temperature of 20 °C, together with the gain medium and pump specifications described previously. The crystal temperature distribution at the center point along the thickness direction is shown in Fig. 3(a). Compared with the result of traditional active mirror module (a single slab with single-sided cooling, that is, the gain medium depicted at the bottom of Fig. 2 with cooling water flow only on its lower face, while all other parameters remain unchanged) under the same heat intensity, which is illustrated in dashed line in Fig. 3(a), our gain module with two crystals and double-sided immersion cooling has the maximum temperature dropped from 140 °C to 60 °C, and the maximum temperature difference reduced by 70%. The temperature distribution on the central plane along the thickness direction of each slab is depicted in Fig. 3(b), showing a cylindrical-shaped thermal lensing, due to the beam folding along the x direction within the active mirror.

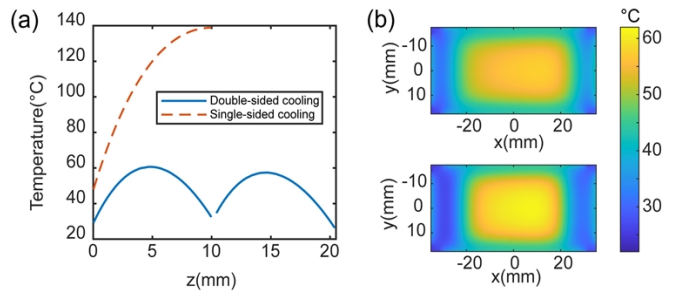

Fig. 3. Simulated temperature distribution: (a) Temperature profile at the center point along the slab thickness; (b) Temperature distribution on the thickness-wise central plane of each slab in the same gain module, where the lower subfigure shows the slab nearer to the pump source.

The small-signal gain of the module is measured to be 4.5 under a pump energy of 18 J. Based on the formula for stored energy, $E_{sto} = g_0 E_s V$, where $g_0$ is the small-signal gain coefficient, $E_s$ is the saturation fluence, and $V$ is the gain volume, the calculated stored energy $E_{sto}$ of each gain module is 6.6 J. At a total pump energy of 90 J, the liquid-immersion-cooled amplifier produces an output energy of 11 J at the repetition rate of 100 Hz and the pulse duration of 7 ns, corresponding to an optical-optical efficiency of 12.2% and the stored energy extraction efficiency of 33.0%. Figure 4 shows the energy stability over 32,000 pulses in 320 s with a root-mean-square (RMS) value of 0.4%, indicating the capability of stable operation. To the best of our knowledge, this result represents the highest energy ever reported for room-temperature nanosecond lasers operating above 100 Hz.

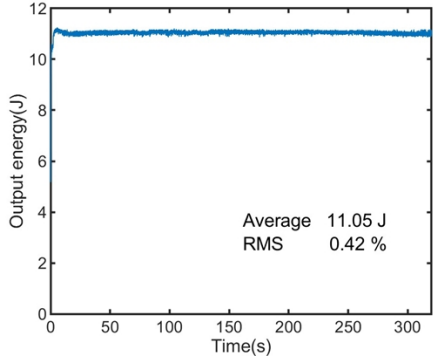

Fig. 4. 320-second temporal stability of output energy at the repetition rate of 100 Hz.

As a result, the measured beam wavefront of the amplified output by a Shack-Hartmann sensor, as shown in Fig. 5(a), indicates that the primary wavefront distortions are defocus and astigmatism, with a noticeable lateral cylindrical wavefront. By optimizing the positions of the spherical and cylindrical lenses within the pre-correction setup, these low-order aberrations are effectively compensated. The resulting corrected wavefront is shown in Fig. 5(b), where the RMS value is dramatically reduced from 5.94 λ to 0.60 λ. The residual wavefront aberrations are then corrected using an AO system incorporating a homemade 116-actuator piezoelectric deformable mirror (DM)[13], as shown in Fig. 5(c), lowering the RMS value down to 0.09 λ. Figure 6(a) and (b) illustrates the beam profiles after wavefront correction in both near field and far field, with the beam quality measured at β = 2.6, which is defined as the ratio of the divergence angle containing 84% of the far-field power (Power in the Bucket, PIB) to that of an ideal plane wave with the same near-field aperture.

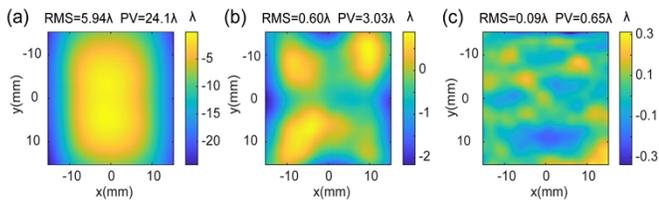

Fig. 5. Measured wavefronts: (a) Thermally induced wavefront aberration of the amplified output; (b) Wavefront after correction by lenses; (c) Wavefront after AO correction.

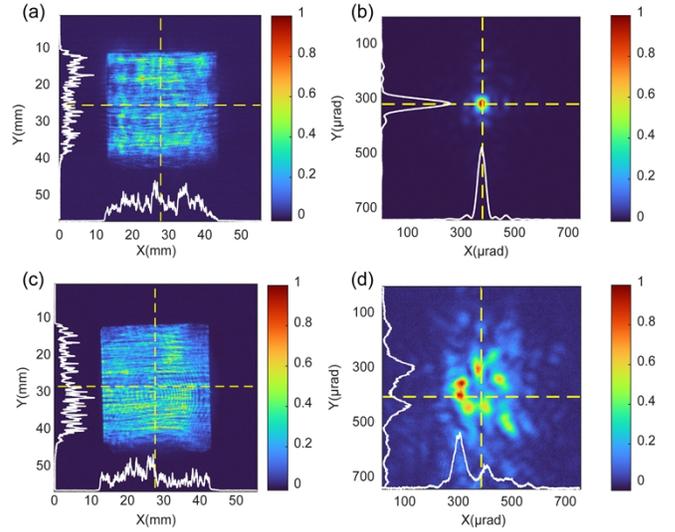

Fig. 6. Beam profiles after wavefront correction: (a) Near field at the flow rate of 5 m/s; (b) Far field at the flow rate of 5 m/s; (c) Near field at the flow rate of 2 m/s; (d) Far field at the flow rate of 2 m/s.

Furthermore, we investigated the influence of cooling flow rate on the corrected beam quality. As the flow rate reduced from 5 m/s to 2 m/s, the beam profile in the far field is severely deteriorated, leading to the beam quality β above 10 even with AO, as shown in Fig. 6(c) and (d). It is because that at the reduced flow rate, the flow-induced wavefront distortions contain larger amplitudes at higher spatial frequencies that exceed the correction bandwidth of the AO system. Thus, it is crucial to realize a high cooling flow rate to maintain the good beam quality for high-repetition-rate operation of water immersion-cooled lasers.

In conclusion, we have demonstrated a diode-pumped Nd:YAG laser system using water immersion cooling, achieving an output energy of 11 J at room temperature at the repetition rate of 100 Hz and the pulse duration of 7 ns, while the beam quality is controlled within 2.6 times diffraction limit. The system maintained stable output with no measurable decline or laser-induced damage over a continuous 320-s test, confirming effective thermal management under room-temperature operation. This result provides a foundation for future engineering toward sustained operation for many hours, as required for practical applications (e.g., the 45-minute run time demonstrated in a cryogenic system [5]). Critically, room-temperature operation eliminates the need for complex and bulky cryogenic cooling systems that are often confined to laboratory settings. This enables the development of more compact and potentially mobile high-energy laser systems. This work demonstrates the great potentials of active mirror nanosecond pulse lasers by liquid immersion cooling for achieving high-average-power and high-repetition-rate room-temperature operation.

Building upon this demonstration, several avenues for future research are envisaged. First, direct performance scaling will be pursued, including increasing the output pulse energy through more gain modules and higher-brightness pump sources, and further improving the beam near-field uniformity and polarization purity via wavefront and thermal management. Second, efforts will focus on engineering a compact, robust prototype suitable for field applications. Finally, exploring nonlinear frequency conversion (e.g., second harmonic generation) to access other spectral regions would significantly broaden the potential applications of this high-energy, high-repetition-rate source.

**Funding.** National Key Research and Development Program of China (No. 2022YFB3606300, No. 2023YFB3610901), Tsinghua University (Department of Precision Instrument)-North Laser Research Institute Co., Ltd Joint Research Center for Advanced Laser Technology (20244910194).

**Disclosures.** The authors declare no conflicts of interest.